\documentclass[twocolumn]{aastex701}
\usepackage{orcidlink}
\usepackage{graphicx}
\usepackage{amsmath}
\newcommand{\teff}{$T_{\!\mbox{\scriptsize\it eff}}$}	% equal to tefftext
\newcommand{\logg}{log\,$g$}
\newcommand{\loggf}{log\,$g_{\!\mbox{\scriptsize\it F}}$}

\newcommand{\one}{\sc i\rm}

\newcommand{\bsg}{\mbox{\small BSG}}

\shorttitle{Short title: Tully Fisher relationship}
\shortauthors{Kudritzki et al.}

\begin{document}

\title{Blue supergiants and the zero point of the Tully-Fisher relation: a path to a new independent test of the Hubble constant}

\author{Rolf-Peter Kudritzki}
\affiliation{Institute for Astronomy, University of Hawaii, 2680 Woodlawn Drive, Honolulu, HI 96822, USA}
\affil{Universit\"ats-Sternwarte, Fakult\"at f\"ur Physik, Ludwig-Maximilians Universit\"at M\"unchen, Scheinerstr. 1, D-81679 M\"unchen, Germany}
\email[]{kud@ifa.hawaii.edu} 

\author[orcid=0000-0002-5068-9833]{Fabio Bresolin}
\affiliation{Institute for Astronomy, University of Hawaii, 2680 Woodlawn Drive, Honolulu, HI 96822, USA}
\email[]{bresolin@ifa.hawaii.edu}  

\author[orcid=0000-0002-9424-0501]{Miguel A. Urbaneja}
\affil{Universit\"at Innsbruck, Institut f\"ur Astro- und Teilchenphysik\\ 
	Technikerstr. 25/8, 6020 Innsbruck, Austria}
\email[]{Miguel.Urbaneja-Perez@uibk.ac.at}  

\author[orcid=0009-0001-5618-4326]{Eva Sextl}
\affil{Universit\"ats-Sternwarte, Fakult\"at f\"ur Physik, Ludwig-Maximilians Universit\"at M\"unchen, Scheinerstr. 1, D-81679 M\"unchen, Germany}
\email[]{sextl@usm.lmu.de}

\correspondingauthor{Rolf-Peter Kudritzki}
\email{kud@ifa.hawaii.edu}
%% A significant change from AASTeX v6+ is in the author blocks. Now an email
%% address is required for each author. This means that each author requires
%% at least one of the following:
%%
%% \author
%% \affiliation
%% \email
%%

\begin{abstract}

Blue supergiant distances of nearby galaxies obtained with the flux-weighted gravity--luminosity relationship are used for a measurement of the zero points of Tully-Fisher relationships at different photometric passbands. The Cousins I-band and the infrared WISE bands W1 and W2 are investigated. The results are compared with previous work using Cepheid and Tip-of-the-Red-Giant-Branch distances. No significant differences were encountered. This supports the large values of the Hubble constant greater than 73~km\,s$^{-1}$\,Mpc$^{-1}$ found with the Tully-Fisher distance ladder work over the last decade. Applying blue supergiant distances on the I-band Tully-Fisher relation observations yields a Hubble constant $H_0 = 76.2\pm6.2$ km\,s$^{-1}$\,Mpc$^{-1}$. The large uncertainty is caused by the still relatively small blue supergiant galaxies sample size but will be reduced in future work.

\end{abstract}



%%%%%%%%%%%%%%%%%%%%%%%%%%%%%%%%%%%%%%%%%%%%%%%%%%%%%%%%%%%%%%%%%%%%%%%%%%%%%%%%%%%%%%%%%%%%%%%%%%%%%%%%%%%%%%%%%%%%%%%%%%%%%%%%%%%%%%%%%%%
% Main Paper

\section{Introduction}\label{sec:intro}
The Hubble tension describes a 5$\sigma$ discrepancy between measurements of the Hubble constant $H_0$ using the distance ladder approach with Cepheid or Tip-of-the-Red-Giant-Branch (TRGB) stars and supernovae of type Ia (\citealt{Riess:2022}, $H_0 = 73.04\pm1.04$ km\,s$^{-1}$\,Mpc$^{-1}$) and the analysis of the angular scales of the cosmic microwave background (CMB) based on the $\Lambda$CDM model of the expanding universe (\citealt{Planck:2020}, $H_0 = 67.04\pm0.5$ km\,s$^{-1}$\,Mpc$^{-1}$). If real, this discrepancy suggests new physics in addition to the $\Lambda$CDM standard model \citep{DiValentino:2021,Schoeneberg:2022}. Recent distance ladder work with JWST \citep{Riess:2024} on the one hand and the measurement of baryon acoustic oscillations (BAO) with the Dark Energy Spectroscopic Instrument (DESI, \citealt{Adame:2025}) on the other have confirmed the discrepancy. However, at the same time distance ladder investigations within the Chicago-Carnegie Hubble Program (CCHP) find agreement with the large-scale structure results \citep{Freedman:2025, Lee:2025}. In this situation, the use of alternative methods becomes important.

The application of the Tully-Fisher relationship (TFR), which uses the correlation between galaxy luminosity and rotation rate \citep{Tully:1977} instead of the peak brightness of SN~Ia for the last rung of the ladder, is such an alternative. This method is well established and has been continuously improved over the past years. Most recently, \cite{Scolnic2024} in their TFR determination of $H_0$ based on the work by \cite{Boubel2024} and using the Cosmicflows-4 database \citep{Kourkchi:2020a,Kourkchi:2020b,Kourkchi:2022} have pointed out that practically all TFR applications over the past decade have resulted in values of $H_0$ larger than 73~km\,s$^{-1}$\,Mpc$^{-1}$, providing clear additional evidence for the Hubble tension. 

However, in all the work carried out over the past few years the TFR zero points have been determined using Cepheid- or TRGB-based distances of nearby galaxies. The question arises whether the use of these stellar distance calibrators is responsible for the large values of $H_0$ found and whether the use of different stellar distance indicators would result in smaller values. In this paper we make a first step to address this question by introducing an alternative, well established accurate distance indicator for the calibration of the TFR zero point: the flux-weighted gravity--luminosity relationship (FGLR) of blue supergiant stars. The number of TFR galaxies with accurate FGLR distances is still small. Thus, the goal of this first step is not to come up with the ultimate new zero point calibration but rather to test how the FGLR method combined with the TFR works and to obtain a first impression on which direction the new distance indicator will lead us.

Our paper has a simple structure. First, we introduce the FGLR method. Then we discuss the TFR in the I-band and the WISE W1- and W2-bands and provide new zero points based on FGLR distances. In an additional brief separate section we consider the effects of TFR curvature. Finally, in the last section, we use the BSG-based I-band zero point to provide a new independent estimate of the Hubble constant and discuss the potential of future extended extragalactic BSG spectroscopy.

% - - - - - - - - - - - - - - - - - - - - - - - - - - - - - - - - - - - - - - - - -
\begin{figure}[ht]
	\center \includegraphics[width=1\columnwidth]{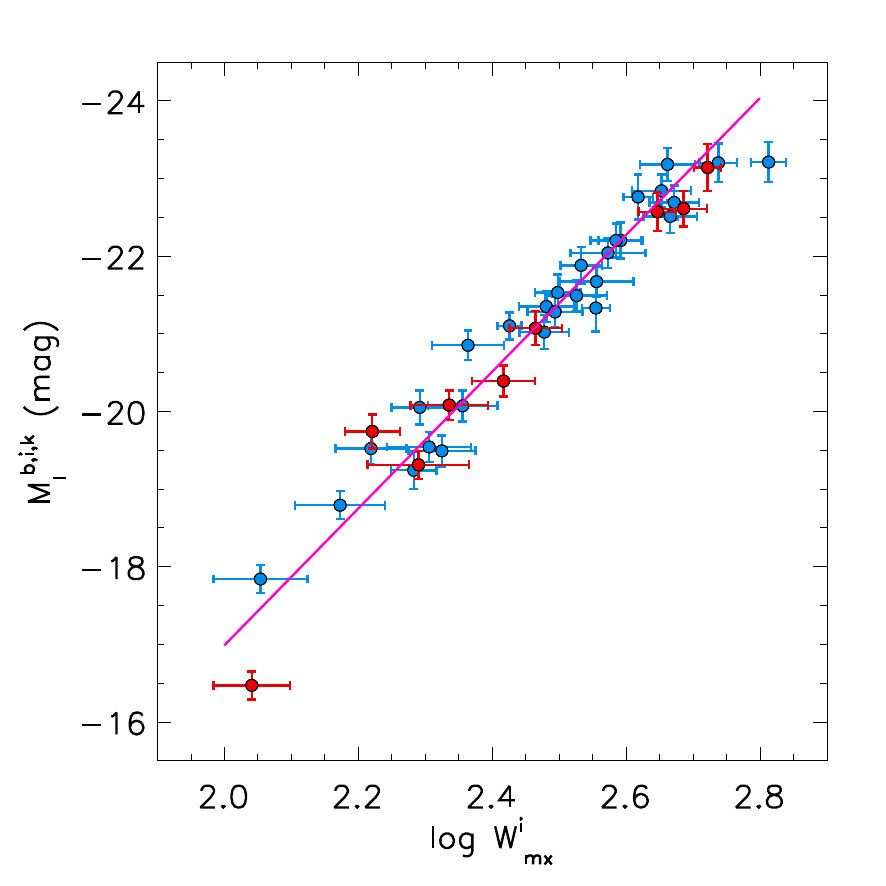}\medskip
	\caption{I-band TFR of the \cite{Tully:2012} zero point calibration sample. The galaxies shown in red are the subsample for which we have determined FGLR distances.}\label{fig:Iband_all}
\end{figure}
% - - - - - - - - - - - - - - - - - - - - - - - - - - - - - - - - - - - - - - - - -

% - - - - - - - - - - - - - - - - - - - - - - - - - - - - - - - - - - - - - - - - -
\begin{figure}[ht]
	\center \includegraphics[width=1\columnwidth]{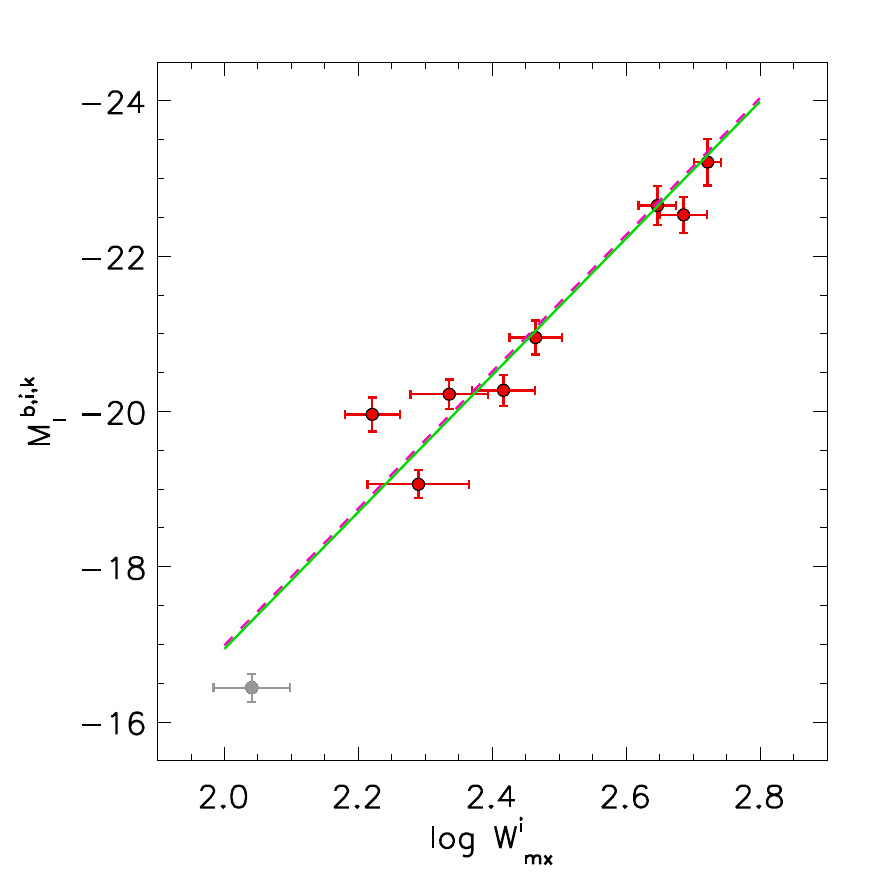}\medskip
	\caption{I-band TFR of the FGLR galaxies using FGLR distance moduli for the calculation of absolute magnitudes. The green line corresponds to the zero point obtained with these galaxies. The pink dashed line is the old regression from Figure \ref{fig:Iband_all}. The galaxy in gray is NGC~3109. As in the work by \cite{Tully:2012} it is not used for the zero point fit.}\label{fig:FGLR_Iband}
\end{figure}
% - - - - - - - - - - - - - - - - - - - - - - - - - - - - - - - - - - - - - - - - -

%==================================================================================
\begin{deluxetable*}{lcccc}
	\tabletypesize{\footnotesize}	
	\tablecaption{FGLR galaxies and I-band Tully-Fisher parameters\label{table:1}}
	
	\tablehead{
		%\colhead{\phn galaxy \phn}	     		&
        \colhead{Galaxy}	     		&
        \colhead{$\log W_{mx}^{i}$}               &
		\colhead{$I^{b,i,k}_{I}$}	 			&
		\colhead{$(m-M)_{TC}$}	 			    &
		\colhead{$(m-M)_{FGLR}$}\\[-2.ex]
		\colhead{}       		&
        \colhead{(km\,s$^{-1}$)}          &
		\colhead{(mag)}       	&
		\colhead{(mag)}       	&
		\colhead{(mag)}\\[-5ex] }

	\colnumbers
	\startdata
	\\[-4.5ex]
	NGC~55   & $2.221\pm0.041$ & $6.88\pm0.22$ & $26.62$ & $26.84\pm0.07$ \\[-0.ex]
M31     & $2.722\pm0.020$ & $1.34\pm0.30$ & $24.48$ & $24.55\pm0.07$ \\[-0.ex]
NGC~300  & $2.290\pm0.076$ & $7.28\pm0.18$ & $26.59$ & $26.34\pm0.06$ \\[-0.ex]
M33     & $2.336\pm0.058$ & $4.74\pm0.19$ & $24.82$ & $24.96\pm0.05$ \\[-0.ex]
NGC~2403 & $2.417\pm0.047$ & $7.11\pm0.20$ & $27.50$ & $27.38\pm0.08$ \\[-0.ex]
M81     & $2.686\pm0.035$ & $5.20\pm0.23$ & $27.81$ & $27.73\pm0.04$ \\[-0.ex]
NGC~3621 & $2.465\pm0.039$ & $8.01\pm0.22$ & $29.08$ & $28.96\pm0.14$ \\[-0.ex]
NGC~4258 & $2.647\pm0.028$ & $6.84\pm0.25$ & $29.41$ & $29.49\pm0.07$ \\[-0.ex]
NGC~3109 & $2.041\pm0.057$ & $9.15\pm0.18$ & $25.62$ & $25.59\pm0.04$ \\[-0.ex]
	\\[-2.5ex]
	\enddata
	\tablecomments{NGC~3109 is not used for the zero point fit. }
\end{deluxetable*}

%==================================================================================

\section{The FGLR distance determination method}\label{sec:FGLR}

Quantitative stellar spectroscopy allows the determination of stellar effective temperature \teff, gravity log g, and metallicity [Z] = log Z/Z$_\odot$. As has been shown by \cite{Kudritzki:2003,Kudritzki:2008,Kudritzki:2020}, the flux-weighted gravity defined as \loggf~= \logg$-$4log(\teff/10$^4$K) is tightly correlated with stellar luminosity and the resulting flux-weighted--gravity luminosity relationship (FGLR) can be used to determine accurate stellar distances. The ideal targets for the extragalactic application of the FGLR are blue supergiant stars (\bsg), i.e.~massive stars in the mass range between 20 to 50 M$_{\odot}$, that cross the Hertzsprung-Russell diagram from the hot main sequence to the cool red supergiant phase. \bsg\ are the brightest stars in the universe at visual light with absolute magnitudes up M$_V$ = $-9$~mag. As such, they can easily be identified in star-forming galaxies as bright blue point sources and multi-object low resolution spectroscopy with telescopes such as Keck or the ESO VLT, in conjunction with the use of non-LTE model atmospheres, then yields stellar temperatures, gravities and metallicities. 

For the BSG spectral analysis a dense grid of state of the art non-LTE model atmosphere spectra is used \citep{Przybilla:2006, Kudritzki:2008, Kudritzki:2012, Nieva:2012} with effective temperatures from 7900 to 15,000 K and gravities log g ranging between 0.8 and 3.0 dex (in cgs units). The metallicities adopted for the model spectra range from a factor of 30 below to a factor of 3 above the solar value. The analysis method is described in detail in \cite{Kudritzki:2024} and \cite{Bresolin:2025} and references therein. It has been carefully tested with BSG in the solar neighborhood \citep{Przybilla:2006, Nieva:2012} and the LMC \citep{Urbaneja:2017} and by a comparison with red supergiant stars, which were analyzed by means of a completely independent method \citep{Gazak:2015}.

The FGLR has been calibrated in the LMC by \cite{Urbaneja:2017}, who carried out a detailed analysis of 90 BSG and obtained a tightly constrained FGLR. The relationship has then been updated by \cite{Sextl:2021} (see their Figure 1) using the one percent precision distance to the LMC measured by \cite{Pietrzynski:2019}. For the spectral analysis of BSG in galaxies beyond the LMC we use the same set of model atmospheres. Thus, the FGLR method works strictly differentially relative the LMC.

We have a long-standing program to investigate nearby galaxies, and the study of \bsg\ has provided important information about chemical evolution of galaxies, interstellar extinction, and accurate distances. Examples of the most recent work are the study of the Hubble constant anchor galaxy NGC 4258 \citep{Kudritzki:2024} and the Pinwheel galaxy M101 \citep{Bresolin:2025}. This work has confirmed the reliability of the FGLR method as a precise distance indicator and has also led to an independent estimate of the Hubble constant by using the peak brightness of a Type~Ia supernova.

We note that the FGLR is a method that is completely independent of the well established standard methods using the Cepheid period luminosity relationship or the TRGB. The advantage of the FGLR method is that it allows to directly and accurately constrain interstellar extinction for each observed individual supergiant target through the direct spectroscopic determination of stellar temperature, gravity, and metallicity. With these stellar parameters the intrinsic supergiant spectral energy distributions are well known and a comparison with the observed photometry leads to a precise determination of reddening and extinction, which is crucial for distance determinations. 

A further advantage of the FGLR method is that BSG are several magnitudes brighter than Cepheid or TRGB stars. Therefore, blending with other objects is much less important. In addition, since we have a spectrum available for each BSG target, significant light contributions by other sources can usually be identified spectroscopically. Those contaminated targets are then excluded.

% - - - - - - - - - - - - - - - - - - - - - - - - - - - - - - - - - - - - - - - - -
\begin{figure}[ht]
	\center \includegraphics[width=1\columnwidth]{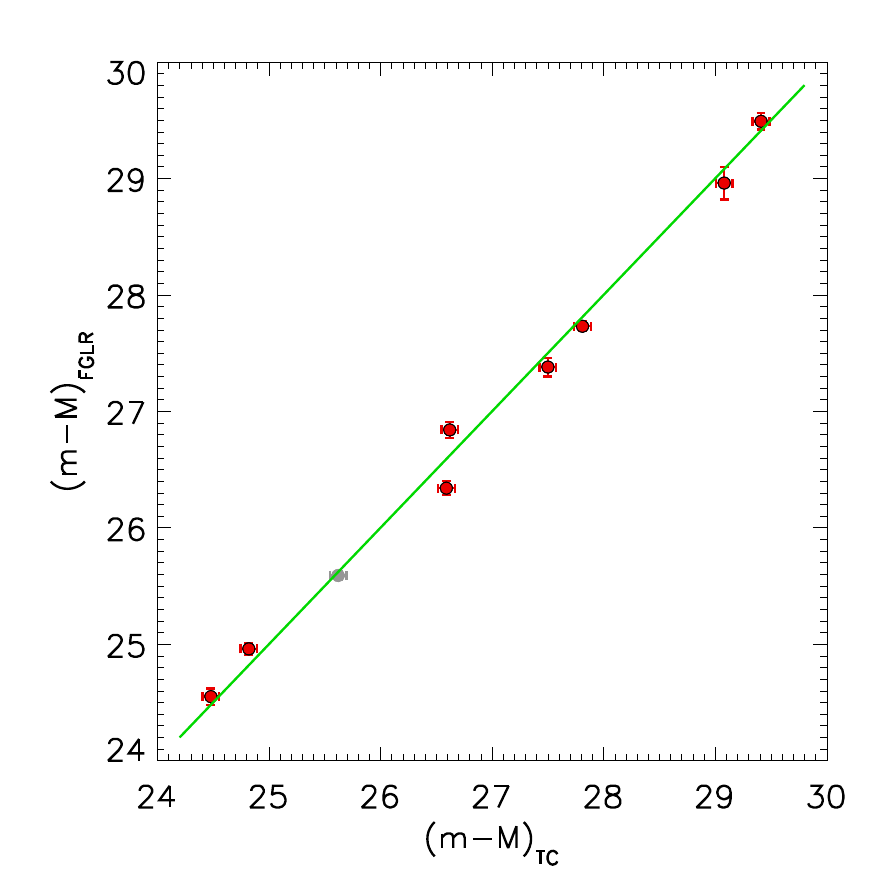}\medskip
	\caption{Distance moduli obtained with the FGLR method versus the distance moduli used in \cite{Tully:2012}.}\label{fig:distmod}
\end{figure}
% - - - - - - - - - - - - - - - - - - - - - - - - - - - - - - - - - - - - - - - - -

\section{The I-band TFR}\label{sec:Iband}

We start our work with the TFR in the Cousins I-band. \citet[hereafter TC12]{Tully:2012} have compiled a sample of 267 calibrator galaxies in 13 galaxy clusters with inclination-corrected 21cm line widths representing galaxy rotation and with Cousins I-band apparent magnitudes corrected for interstellar extinction and also inclination. These galaxies are used to determine the slope of the TFR. The zero point is then determined from a set of 36 nearby galaxies also with corrected I-band photometry and line widths, but in addition with distances obtained from Cepheid and/or TRGB distance determinations. The resulting TFR is

\begin{equation}
    M^{b,i,k}_I = -21.39 - 8.81(\log W^i_{mx} -2.5),
\end{equation}

where $M^{b,i,k}_I$ is the corrected absolute I-band magnitude in the Vega system and $W^i_{mx}$ the corrected line width measured in km/s. Figure \ref{fig:Iband_all} shows the TC12 zero point calibration sample. The fit relation of Equation~(1) is shown in pink. Data are obtained from Table~2 in TC12 and Table~1 in \cite{Neill:2014}. The two galaxies with $\log W_{mx}^{i} \le 2.1$ were not included in the determination of the zero point. The scatter around the relation is $\sigma = 0.36$ mag.

A subsample of the galaxies displayed in Figure~\ref{fig:Iband_all} has been observed in the course of our \bsg\ spectroscopy program. These galaxies are plotted in red, and the necessary information about them is provided in Table~\ref{table:1}. $I^{b,i,k}_{I}$ is the apparent I-band magnitude and $(m-M)_{TC}$ the distance modulus adopted by TC12. The distance moduli obtained with our FGLR distance determination method are given in the last column of the table as $(m-M)_{FGLR}$. They are taken from Table~3 in \cite{Bresolin:2025}. We note that not all galaxies for which we have determined FGLR distances are included in our Table~\ref{table:1}: M101 and M81 are almost face on spirals and IC~1613 and WLM are irregular dwarf galaxies.

Figure \ref{fig:FGLR_Iband} shows the TFR of the FGLR galaxies with absolute magnitudes obtained from the application of the FGLR distance moduli. Adopting the same slope as in Equation~(1) we can then determine a new zero point using the FGLR galaxies. We obtain $-21.344$ mag, a value very close to the TC12 zero point but with a larger uncertainty ($\pm$0.17 mag) due to the smaller sample size. The scatter is $\sigma = 0.50$ mag. The corresponding TFR is shown in Figure~\ref{fig:FGLR_Iband} as a green line. The original TC12 relationship is plotted as a pink dashed line. The difference is very small. We note that as in TC12 the low luminosity galaxy NGC~3109 is not included in the zero point estimate. TC12 argued that compared to the many other galaxies used for their determination of the TFR slope, NGC~3109 is too faint and thus an outlier that should be excluded.  

It is important to test how much of the small difference between the TC12 and the FGLR zero points is a selection effect. For this purpose, we determine a zero point from our eight FGLR galaxies but with the TC12 distances. We obtain $-21.352$ mag, a value very close to the FGLR distance result. Thus, a major part of the difference results from the fact that most of the FGLR galaxies are below the regression curve in Figure~\ref{fig:FGLR_Iband}. The difference between the TC12 and FGLR distance moduli is almost negligible. This is confirmed by Figure~\ref{fig:distmod}, where the distance moduli are compared (note that we adopted 0.07 mag uncertainties for the plot of the TC12 data. No errors are given in TC12). The mean value of the difference is 0.008 mag, exactly the difference between the TC12 and FGLR zero points of the 8 galaxies of the FGLR sample. This is a strong confirmation of the Cepheid/TRGB-based TFR zero points with a new independent distance determination method.   \\

%==================================================================================
\begin{deluxetable*}{lccc}
	\tabletypesize{\footnotesize}	
	\tablecaption{FGLR galaxies and WISE W1-, W2-band Tully-Fisher parameters\label{table:2}}
	
	\tablehead{
		%\colhead{\phn galaxy \phn}	     		&
        \colhead{Galaxy}	     		&
        \colhead{$\log W_{mx}^{i}$}               &
		\colhead{$m^{b,i,k}_{W1}$}	 			&
		\colhead{$m^{b,i,k}_{W2}$}\\[-2.ex]
		\colhead{}       		&
        \colhead{(km\,s$^{-1}$)}          &
		\colhead{(mag)}       	&
		\colhead{(mag)}\\[-5ex] }

	\colnumbers
	\startdata
	\\[-4.5ex]
	NGC~55     & $2.222\pm0.034$ &   $8.04\pm0.05$  &  $8.60\pm0.05$  \\[-0.ex]
NGC~300   & $2.244\pm0.063$ &   $8.25\pm0.05$  &  $8.87\pm0.05$  \\[-0.ex]
M33         & $2.319\pm0.023$ &   $5.62\pm0.05$  &  $6.47\pm0.05$  \\[-0.ex]
NGC~2403 & $2.454\pm0.026$ &   $8.26\pm0.05$  &  $8.79\pm0.05$  \\[-0.ex]
M81         & $2.706\pm0.020$ &   $6.24\pm0.05$  &  $6.89\pm0.05$  \\[-0.ex]
NGC~3621 & $2.468\pm0.026$ &   $8.89\pm0.05$  &  $9.48\pm0.05$  \\[-0.ex]
NGC~4258 & $2.652\pm0.013$ &   $7.86\pm0.05$  &  $8.47\pm0.05$  \\[-0.ex]
NGC~3109 & $2.046\pm0.051$ & $10.40\pm0.05$  & $11.03\pm0.05$ \\[-0.ex]
	\\[-2.5ex]
	\enddata
	\tablecomments{NGC~3109 is not used for the zero point fit. }
\end{deluxetable*}

%==================================================================================

\begin{figure}[ht]
	\center \includegraphics[width=1\columnwidth]{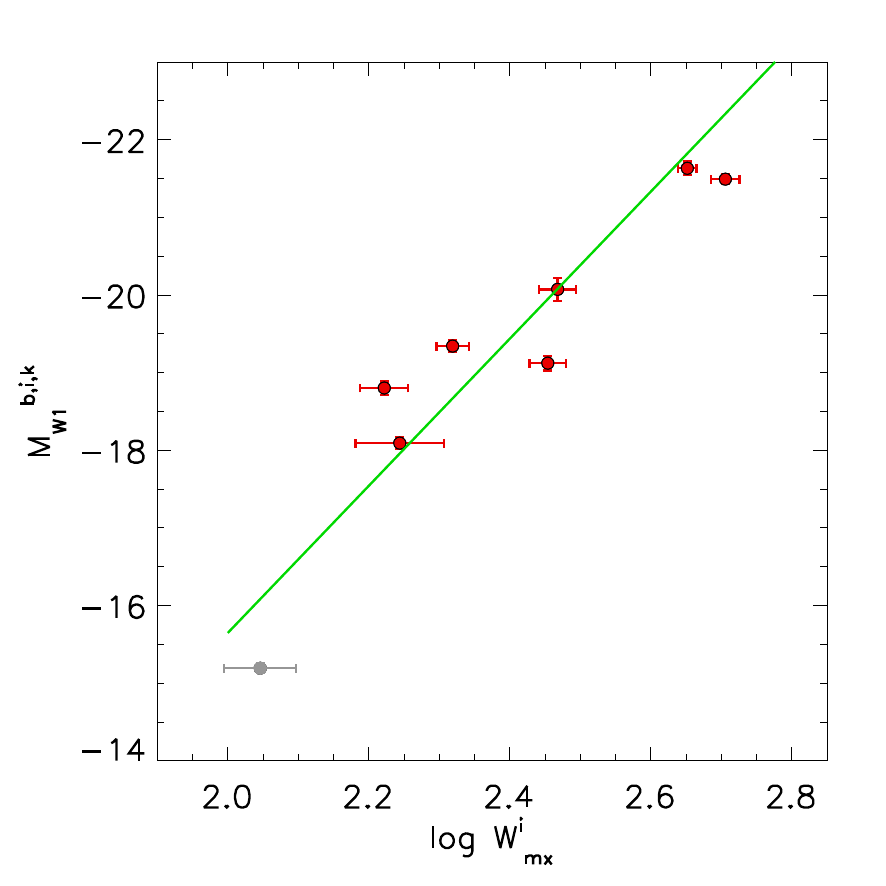}\medskip
	\caption{WISE W1-band Tully-Fisher relation of the FGLR galaxies using FGLR distance moduli for the calculation of absolute magnitude. The green line corresponds to the zero point obtained with these galaxies. The galaxy in gray is NGC~3109. It is not used for the zero point fit.}\label{fig:FGLR_W1band}
\end{figure}
% - - - - - - - - - - - - - - - - - - - - - - - - - - - - - - - - - - - - - - - - -

\begin{figure}[ht]
	\center \includegraphics[width=1\columnwidth]{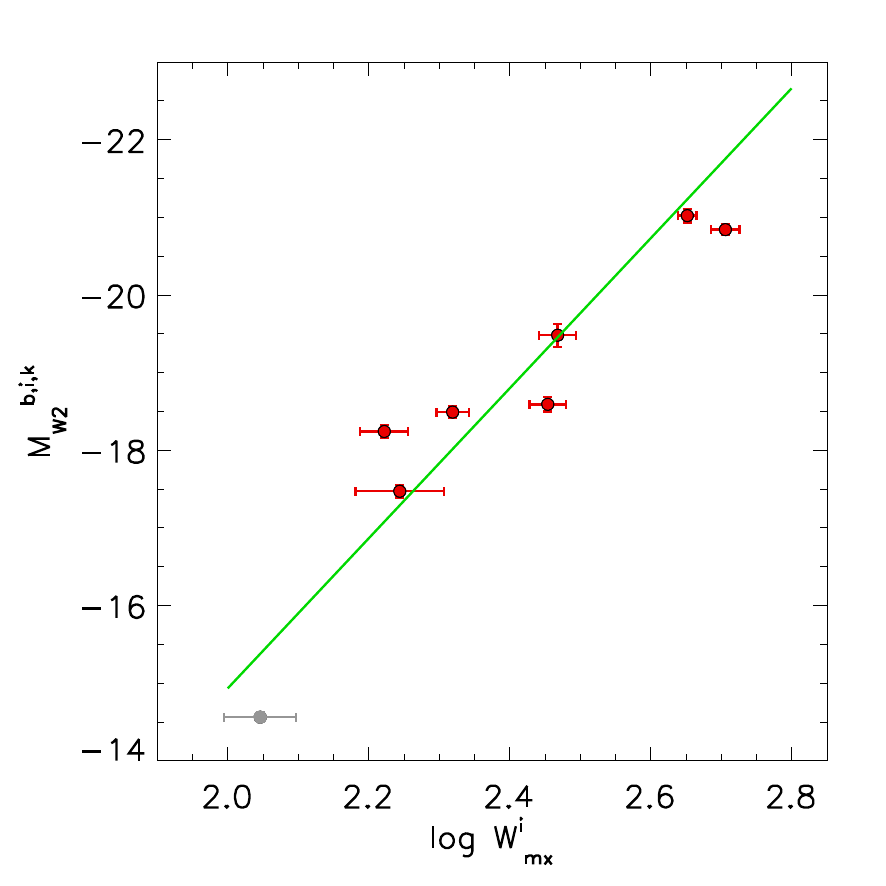}\medskip
	\caption{WISE W2-band Tully-Fisher relation of the FGLR galaxies using FGLR distance moduli for the calculation of absolute magnitude. The green line corresponds to the zero point obtained with these galaxies. The galaxy in gray is NGC~3109. It is not used for the zero point fit.}\label{fig:FGLR_W2band}
\end{figure}
% - - - - - - - - - - - - - - - - - - - - - - - - - - - - - - - - - - - - - - - - -

\section{WISE W1 and W2 TFR}\label{sec:WISE}

\citet[hereafter K20]{Kourkchi:2020a} significantly extended the work on the TFR by establishing and then using the Cosmicflows-4 database. In addition to considering more galaxies they also revised the HI line width measurements, the photometry and Cepheid/TRGB distances for the zero point calibrator galaxies. They studied the relationship at many wavelengths, from the near ultraviolet to the infrared. The infrared relationships are based on photometry obtained with the Wide Field Infrared Survey Explorer (WISE) in the passbands W1 and W2.  The resulting TFR were

\begin{equation}
    M^{b,i,k}_{W1} = -20.36 - 9.47(\log W^i_{mx} -2.5),
\end{equation}
\begin{equation}
    M^{b,i,k}_{W2} = -19.76 - 9.66(\log W^i_{mx} -2.5).
\end{equation}

$M^{b,i,k}_{W1}$ and $M^{b,i,k}_{W2}$ are the extinction- and inclination- corrected absolute WISE magnitudes in passbands W1 and W2, respectively. They are in the AB system. The calibration procedure to obtain these relations was similar to TC12. The slopes of the relation were obtained from $\sim$600 spiral galaxies in 20 galaxy clusters and the zero points from 64 nearby galaxies with Cepheid and/or TRGB distances.

Compared to this relatively large number of zero point calibrator galaxies our set of galaxies with FGLR distances overlapping with the K20 sample is small. Still, it is tempting to estimate zero points from this sample. 

The information about our FGLR galaxies with WISE photometry measurements is given in Table~\ref{table:2}. It contains the new K20 line widths and the WISE W1 and W2 apparent magnitudes. The photometric uncertainty of 0.05 mag was adopted by K20 and is also used by us. The corresponding TFR are shown in Figures \ref{fig:FGLR_W1band} and \ref{fig:FGLR_W2band}. We obtain zero points very close to the K20 results, $-20.38$ mag for W1 and $-19.76$ for W2. While the scatter around the relations is large ($\sigma = 0.70$ mag), the values of the zero points support the results obtained with the calibration using Cepheid and TRGB stars.

\begin{figure}[ht]
	\center \includegraphics[width=1\columnwidth]{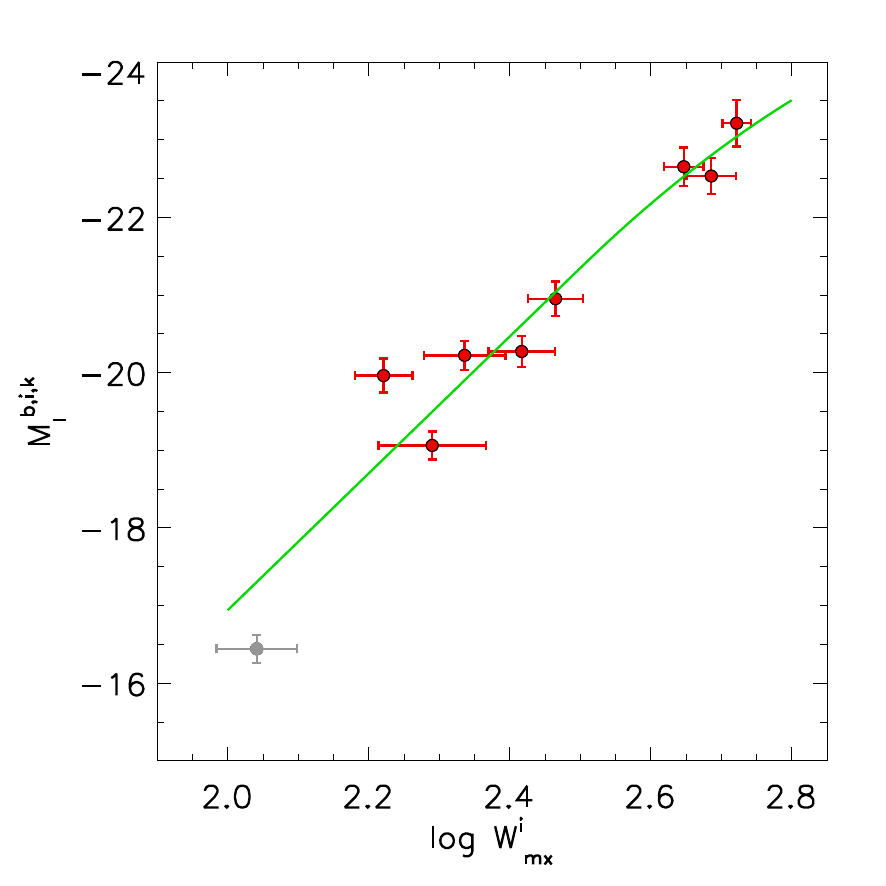}\medskip
	\caption{Curved I-band Tully-Fisher relation of the FGLR galaxies.}\label{fig:CurveI}
\end{figure}

\begin{figure}[ht]
	\center \includegraphics[width=1\columnwidth]{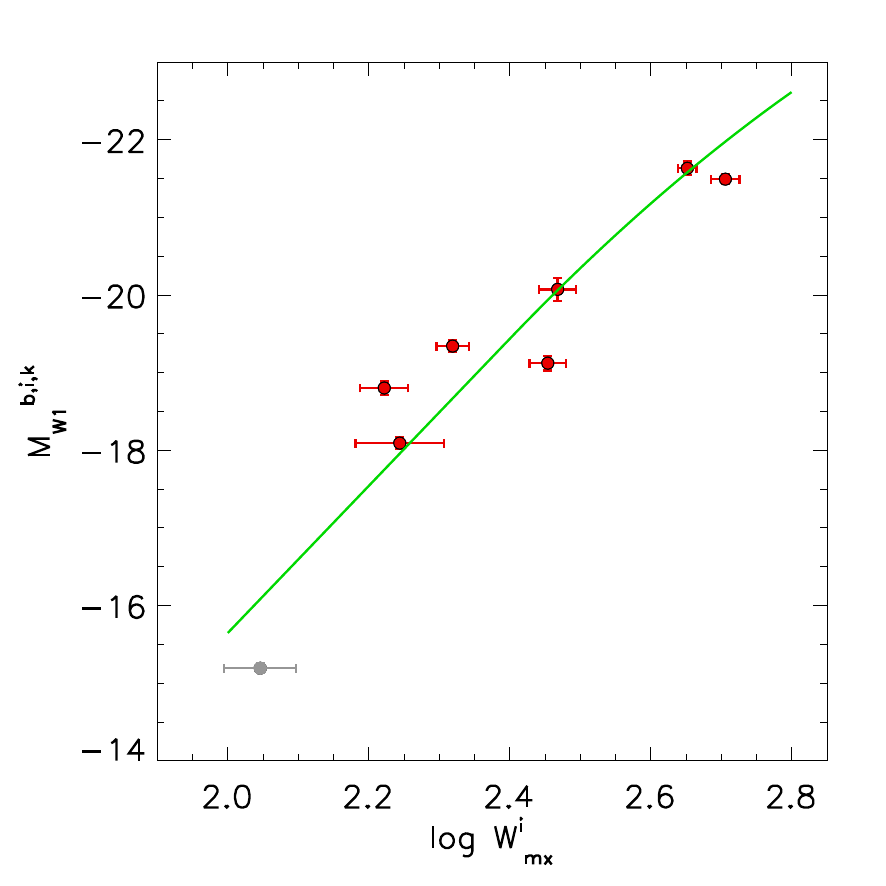}\medskip
    \center \includegraphics[width=1\columnwidth]{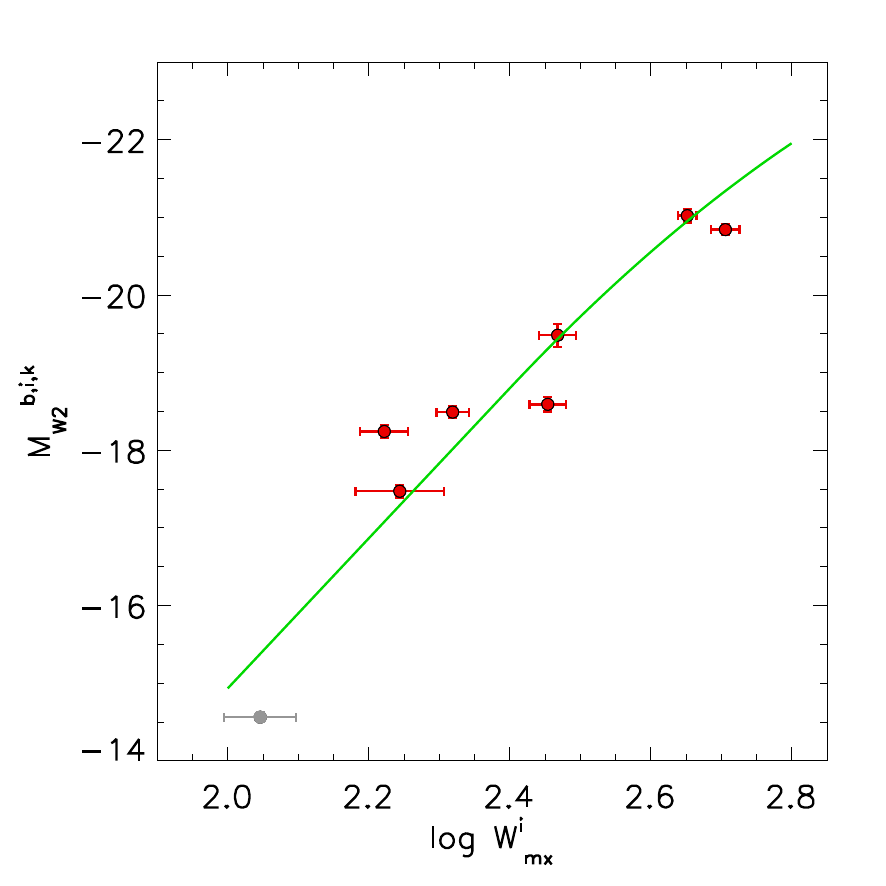}\medskip
	\caption{Curved WISE Tully-Fisher relations of the FGLR galaxies. Top: W1, bottom: W2. }\label{fig:CurveWise}
\end{figure}

\section{TFR curvature}\label{sec:Curv}

K20 and other authors (see, for instance, \citealt{Neill:2014,Boubel2024}) have also discussed the bending of TFR at large HI line widths and introduced curved TFR fits in addition to the linear fits. We follow K20 and introduce a magnitude correction

\begin{equation}
    \Delta M = A_0 + A_1(\log W^i_{mx} -2.5) + A_2(\log W^i_{mx} -2.5)^2
\end{equation}

for 
\begin{equation}
    (\log W^i_{mx} -2.5) \ge x_{break}.
\end{equation}

At $\log W^i_{mx} - 2.5$ = x$_{break}$ and below the correction is zero and at the break point the slope of the correction agrees with the slope of the linear relation. 

Following K20 we adopt x$_{break}$ = 0 and A$_2$ = 5.34 for the I-band. The resulting fit of our FGLR galaxies is clearly improved as shown in Figure~\ref{fig:CurveI}. However, while the fit looks significantly better the reduction of $\sigma$ from 0.50 mag to 0.48 mag is only marginal. This agrees with findings by \cite{Neill:2014} and K20 that curvature effects in the I-band are small. (We note that K20 have actually used the SDSS i-filter rather than the Cousins I-band, but for the effects discussed here this does not create a big difference). 

For the WISE infrared filters we adopt x$_{break} = -0.1$ and A$_2$ = 3.81 and 4.42 for W1 and W2, respectively. These are also the same values as in K20. The corresponding fits are displayed in Figure~\ref{fig:CurveWise} and show again a clear improvement. The scatter is reduced to $\sigma$ = 0.62 mag from the original 0.71 mag.

\section{Conclusions}\label{sec:Concl}

Our determination of the TFR zero points with a small set of galaxies with accurately determined FGLR distances has resulted in values very close to those obtained from the use of Cepheid and/or TRGB stars as distance indicators. For the I-band TFR the maximum difference between the zero point of our FGLR fit and the work by TC12 is $\Delta$ZP = 0.046 mag. Since the change of the zero point is equal to the change of TFR absolute magnitude, $\Delta M^{b,i,k}_I$ = $\Delta$ZP, this translates into a change of the Hubble constant of $\Delta$log($H_0$) = 0.2$\Delta$ZP, when the TFR is used as a distance indicator in the distant universe. TC12 obtained $H_0 = 75.1\pm2.7$ km\,s$^{-1}$\,Mpc$^{-1}$. Applying the FGLR zero point then yields $H_0 = 76.7$ km\,s$^{-1}$\,Mpc$^{-1}$. However, we note that the FGLR zero point still has a relatively large uncertainty (0.17 mag), and thus, the Hubble constant error for the FGLR-based determination is $\pm6.2$~km\,s$^{-1}$.  

In summary, the work presented here indicates that the blue supergiant FGLR distance determination method has the potential to become an important alternative for the calibration of the Tully-Fisher relation and its use for the determination of the Hubble constant. There are about ten galaxies with distances smaller than 10 Mpc and HI-line width measurements which we plan to observe during the next two years within the course of the blue supergiant program with the existing 8 to 10m telescopes. This will already help to reduce the present zero point and Hubble constant uncertainties. With the forthcoming extremely large telescopes we will then have the means to go out to 20 Mpc, which will make a large number of additional galaxies available for a very accurate blue supergiant-based calibration of the relation and a subsequent application to determine $H_0$.

%% Please use the acknowledgment and contribution environments. 
\begin{acknowledgments}
RPK and ES acknowledge support by the Munich Excellence Cluster Origins and the Munich Institute for Astro-, Particle and Biophysics (MIAPbP) both funded by the Deutsche Forschungsgemeinschaft (DFG, German Research Foundation) under Germany's Excellence Strategy EXC-2094 390783311. In addition, ES has been supported by the European Research Councel COMPLEX project under the European Union's Horizon 2020 research and innovation program grant agreement ERC-2029-AdG 882679.
\end{acknowledgments}

%\facilities{HST(STIS),...}

%%%%%%%%%%%%%%%%%%%%%%%%%%%%%%%%%%%%%%%%%%%%%%%%%%%%%%%%%%%%%%%%%%%%%%%%%%%%%%%%%%%%%%%%%%%%%%%%%%%%%%%%%%%%%%%%%%%%%%%%%%%%%%%%%%%%%%%%%%%
% the bibligraphy
\bibliography{paper.bib}{}

\begin{thebibliography}{}
\expandafter\ifx\csname natexlab\endcsname\relax\def\natexlab#1{#1}\fi
\providecommand{\url}[1]{\href{#1}{#1}}
\providecommand{\dodoi}[1]{doi:~\href{http://doi.org/#1}{\nolinkurl{#1}}}
\providecommand{\doeprint}[1]{\href{http://ascl.net/#1}{\nolinkurl{http://ascl.net/#1}}}
\providecommand{\doarXiv}[1]{\href{https://arxiv.org/abs/#1}{\nolinkurl{https://arxiv.org/abs/#1}}}

% type= article
\bibitem[{A.~G. {Adame} {et~al.}(2025){Adame}, {Aguilar}, {Ahlen}, {Alam},
  {Alexander}, {Alvarez}, {Alves}, {Anand}, {Andrade}, {Armengaud}, {Avila},
  {Aviles}, {Awan}, {Bahr-Kalus}, {Bailey}, {Baltay}, {Bault}, {Behera},
  {BenZvi}, {Bera}, {Beutler}, {Bianchi}, {Blake}, {Blum}, {Brieden},
  {Brodzeller}, {Brooks}, {Buckley-Geer}, {Burtin}, {Calderon}, {Canning},
  {Carnero Rosell}, {Cereskaite}, {Cervantes-Cota}, {Chabanier}, {Chaussidon},
  {Chaves-Montero}, {Chen}, {Chen}, {Claybaugh}, {Cole}, {Cuceu}, {Davis},
  {Dawson}, {de la Macorra}, {de Mattia}, {Deiosso}, {Dey}, {Dey}, {Ding},
  {Doel}, {Edelstein}, {Eftekharzadeh}, {Eisenstein}, {Elliott}, {Fagrelius},
  {Fanning}, {Ferraro}, {Ereza}, {Findlay}, {Flaugher}, {Font-Ribera},
  {Forero-S{\'a}nchez}, {Forero-Romero}, {Frenk}, {Garcia-Quintero},
  {Gazta{\~n}aga}, {Gil-Mar{\'\i}n}, {Gontcho a Gontcho}, {Gonzalez-Morales},
  {Gonzalez-Perez}, {Gordon}, {Green}, {Gruen}, {Gsponer}, {Gutierrez}, {Guy},
  {Hadzhiyska}, {Hahn}, {Hanif}, {Herrera-Alcantar}, {Honscheid}, {Howlett},
  {Huterer}, {Ir{\v{s}}i{\v{c}}}, {Ishak}, {Juneau}, {Kara{\c{c}}ayl{\i}},
  {Kehoe}, {Kent}, {Kirkby}, {Kremin}, {Krolewski}, {Lai}, {Lan}, {Landriau},
  {Lang}, {Lasker}, {Le Goff}, {Le Guillou}, {Leauthaud}, {Levi}, {Li},
  {Linder}, {Lodha}, {Magneville}, {Manera}, {Margala}, {Martini}, {Maus},
  {McDonald}, {Medina-Varela}, {Meisner}, {Mena-Fern{\'a}ndez}, {Miquel},
  {Moon}, {Moore}, {Moustakas}, {Mueller}, {Mu{\~n}oz-Guti{\'e}rrez}, {Myers},
  {Nadathur}, {Napolitano}, {Neveux}, {Newman}, {Nguyen}, {Nie}, {Niz},
  {Noriega}, {Padmanabhan}, {Paillas}, {Palanque-Delabrouille}, {Pan},
  {Penmetsa}, {Percival}, {Pieri}, {Pinon}, {Poppett}, {Porredon}, {Prada},
  {P{\'e}rez-Fern{\'a}ndez}, {P{\'e}rez-R{\`a}fols}, {Rabinowitz}, {Raichoor},
  {Ram{\'\i}rez-P{\'e}rez}, {Ramirez-Solano}, {Rashkovetskyi}, {Ravoux},
  {Rezaie}, {Rich}, {Rocher}, {Rockosi}, {Roe}, {Rosado-Marin}, {Ross},
  {Rossi}, {Ruggeri}, {Ruhlmann-Kleider}, {Samushia}, {Sanchez}, {Saulder},
  {Schlafly}, {Schlegel}, {Schubnell}, {Seo}, {Shafieloo}, {Sharples},
  {Silber}, {Slosar}, {Smith}, {Sprayberry}, {Tan}, {Tarl{\'e}}, {Taylor},
  {Trusov}, {Ure{\~n}a-L{\'o}pez}, {Vaisakh}, {Valcin}, {Valdes},
  {Vargas-Maga{\~n}a}, {Verde}, {Walther}, {Wang}, {Wang}, {Weaver},
  {Weaverdyck}, {Wechsler}, {Weinberg}, {White}, {Yu}, {Yu}, {Yuan},
  {Y{\`e}che}, {Zaborowski}, {Zarrouk}, {Zhang}, {Zhao}, {Zhao}, {Zhou}, \&
  {Zhuang}}]{Adame:2025}
{Adame}, A.~G., {Aguilar}, J., {Ahlen}, S., {et~al.} 2025,
  \bibinfo{title}{{DESI 2024 VI: cosmological constraints from the measurements
  of baryon acoustic oscillations},} \jcap, 2025, 021,
  \dodoi{10.1088/1475-7516/2025/02/021}

% type= article
\bibitem[{P. {Boubel} {et~al.}(2024){Boubel}, {Colless}, {Said}, \&
  {Staveley-Smith}}]{Boubel2024}
{Boubel}, P., {Colless}, M., {Said}, K., \& {Staveley-Smith}, L. 2024,
  \bibinfo{title}{{An improved Tully-Fisher estimate of H$_{0}$},} \mnras, 533,
  1550, \dodoi{10.1093/mnras/stae1925}

% type= article
\bibitem[{F. {Bresolin} {et~al.}(2025){Bresolin}, {Kudritzki}, {Urbaneja},
  {Sextl}, \& {Riess}}]{Bresolin:2025}
{Bresolin}, F., {Kudritzki}, R.-P., {Urbaneja}, M.~A., {Sextl}, E., \& {Riess},
  A.~G. 2025, \bibinfo{title}{{Blue supergiants in the Pinwheel Galaxy M101:
  comparison with H II region chemical abundances, spectroscopic distance and
  an independent determination of the Hubble constant},} arXiv e-prints,
  arXiv:2508.11837.
\newblock \doarXiv{2508.11837}

% type= article
\bibitem[{E. {Di Valentino} {et~al.}(2021){Di Valentino}, {Mena}, {Pan},
  {Visinelli}, {Yang}, {Melchiorri}, {Mota}, {Riess}, \&
  {Silk}}]{DiValentino:2021}
{Di Valentino}, E., {Mena}, O., {Pan}, S., {et~al.} 2021, \bibinfo{title}{{In
  the realm of the Hubble tension-a review of solutions},} Classical and
  Quantum Gravity, 38, 153001, \dodoi{10.1088/1361-6382/ac086d}

% type= article
\bibitem[{W.~L. {Freedman} {et~al.}(2025){Freedman}, {Madore}, {Hoyt}, {Jang},
  {Lee}, \& {Owens}}]{Freedman:2025}
{Freedman}, W.~L., {Madore}, B.~F., {Hoyt}, T.~J., {et~al.} 2025,
  \bibinfo{title}{{Status Report on the Chicago-Carnegie Hubble Program (CCHP):
  Measurement of the Hubble Constant Using the Hubble and James Webb Space
  Telescopes},} \apj, 985, 203, \dodoi{10.3847/1538-4357/adce78}

% type= article
\bibitem[{J.~Z. {Gazak} {et~al.}(2015){Gazak}, {Kudritzki}, {Evans}, {Patrick},
  {Davies}, {Bergemann}, {Plez}, {Bresolin}, {Bender}, {Wegner}, {Bonanos}, \&
  {Williams}}]{Gazak:2015}
{Gazak}, J.~Z., {Kudritzki}, R., {Evans}, C., {et~al.} 2015,
  \bibinfo{title}{{Red Supergiants as Cosmic Abundance Probes: The Sculptor
  Galaxy NGC 300},} \apj, 805, 182, \dodoi{10.1088/0004-637X/805/2/182}

% type= article
\bibitem[{E. {Kourkchi} {et~al.}(2020{\natexlab{a}}){Kourkchi}, {Tully},
  {Anand}, {Courtois}, {Dupuy}, {Neill}, {Rizzi}, \&
  {Seibert}}]{Kourkchi:2020a}
{Kourkchi}, E., {Tully}, R.~B., {Anand}, G.~S., {et~al.} 2020{\natexlab{a}},
  \bibinfo{title}{{Cosmicflows-4: The Calibration of Optical and Infrared
  Tully-Fisher Relations},} \apj, 896, 3, \dodoi{10.3847/1538-4357/ab901c}

% type= article
\bibitem[{E. {Kourkchi} {et~al.}(2022){Kourkchi}, {Tully}, {Courtois}, {Dupuy},
  \& {Guinet}}]{Kourkchi:2022}
{Kourkchi}, E., {Tully}, R.~B., {Courtois}, H.~M., {Dupuy}, A., \& {Guinet}, D.
  2022, \bibinfo{title}{{Cosmicflows-4: the baryonic Tully-Fisher relation
  providing 10 000 distances},} \mnras, 511, 6160,
  \dodoi{10.1093/mnras/stac303}

% type= article
\bibitem[{E. {Kourkchi} {et~al.}(2020{\natexlab{b}}){Kourkchi}, {Tully},
  {Eftekharzadeh}, {Llop}, {Courtois}, {Guinet}, {Dupuy}, {Neill}, {Seibert},
  {Andrews}, {Chuang}, {Danesh}, {Gonzalez}, {Holthaus}, {Mokelke}, {Schoen},
  \& {Urasaki}}]{Kourkchi:2020b}
{Kourkchi}, E., {Tully}, R.~B., {Eftekharzadeh}, S., {et~al.}
  2020{\natexlab{b}}, \bibinfo{title}{{Cosmicflows-4: The Catalog of
  {\ensuremath{\sim}}10,000 Tully-Fisher Distances},} \apj, 902, 145,
  \dodoi{10.3847/1538-4357/abb66b}

% type= article
\bibitem[{R.~P. {Kudritzki} {et~al.}(2003){Kudritzki}, {Bresolin}, \&
  {Przybilla}}]{Kudritzki:2003}
{Kudritzki}, R.~P., {Bresolin}, F., \& {Przybilla}, N. 2003, \bibinfo{title}{{A
  New Extragalactic Distance Determination Method Using the Flux-weighted
  Gravity of Late B and Early A Supergiants},} \apjl, 582, L83,
  \dodoi{10.1086/367690}

% type= article
\bibitem[{R.-P. {Kudritzki} {et~al.}(2024){Kudritzki}, {Urbaneja}, {Bresolin},
  {Macri}, {Yuan}, {Li}, {Anand}, \& {Riess}}]{Kudritzki:2024}
{Kudritzki}, R.-P., {Urbaneja}, M.~A., {Bresolin}, F., {et~al.} 2024,
  \bibinfo{title}{{The Hubble Constant Anchor Galaxy NGC 4258: Metallicity and
  Distance from Blue Supergiants},} \apj, 977, 217,
  \dodoi{10.3847/1538-4357/ad9279}

% type= article
\bibitem[{R.-P. {Kudritzki} {et~al.}(2008){Kudritzki}, {Urbaneja}, {Bresolin},
  {Przybilla}, {Gieren}, \& {Pietrzy{\'n}ski}}]{Kudritzki:2008}
{Kudritzki}, R.-P., {Urbaneja}, M.~A., {Bresolin}, F., {et~al.} 2008,
  \bibinfo{title}{{Quantitative Spectroscopy of 24 A Supergiants in the
  Sculptor Galaxy NGC 300: Flux-weighted Gravity-Luminosity Relationship,
  Metallicity, and Metallicity Gradient},} \apj, 681, 269,
  \dodoi{10.1086/588647}

% type= article
\bibitem[{R.-P. {Kudritzki} {et~al.}(2012){Kudritzki}, {Urbaneja}, {Gazak},
  {Bresolin}, {Przybilla}, {Gieren}, \& {Pietrzy{\'n}ski}}]{Kudritzki:2012}
{Kudritzki}, R.-P., {Urbaneja}, M.~A., {Gazak}, Z., {et~al.} 2012,
  \bibinfo{title}{{Quantitative Spectroscopy of Blue Supergiant Stars in the
  Disk of M81: Metallicity, Metallicity Gradient, and Distance},} \apj, 747,
  15, \dodoi{10.1088/0004-637X/747/1/15}

% type= article
\bibitem[{R.-P. {Kudritzki} {et~al.}(2020){Kudritzki}, {Urbaneja}, \&
  {Rix}}]{Kudritzki:2020}
{Kudritzki}, R.-P., {Urbaneja}, M.~A., \& {Rix}, H.-W. 2020, \bibinfo{title}{{A
  Simple Unified Spectroscopic Indicator of Stellar Luminosity: The Extended
  Flux-weighted Gravity-Luminosity Relationship},} \apj, 890, 28,
  \dodoi{10.3847/1538-4357/ab67c3}

% type= article
\bibitem[{A.~J. {Lee} {et~al.}(2025){Lee}, {Freedman}, {Madore}, {Jang},
  {Owens}, \& {Hoyt}}]{Lee:2025}
{Lee}, A.~J., {Freedman}, W.~L., {Madore}, B.~F., {et~al.} 2025,
  \bibinfo{title}{{The Chicago{\textendash}Carnegie Hubble Program: The JWST
  J-region Asymptotic Giant Branch Extragalactic Distance Scale},} \apj, 985,
  182, \dodoi{10.3847/1538-4357/adc8a1}

% type= article
\bibitem[{J.~D. {Neill} {et~al.}(2014){Neill}, {Seibert}, {Tully}, {Courtois},
  {Sorce}, {Jarrett}, {Scowcroft}, \& {Masci}}]{Neill:2014}
{Neill}, J.~D., {Seibert}, M., {Tully}, R.~B., {et~al.} 2014,
  \bibinfo{title}{{The Calibration of the WISE W1 and W2 Tully-Fisher
  Relation},} \apj, 792, 129, \dodoi{10.1088/0004-637X/792/2/129}

% type= article
\bibitem[{M.-F. {Nieva} \& N. {Przybilla}(2012){Nieva} \&
  {Przybilla}}]{Nieva:2012}
{Nieva}, M.-F., \& {Przybilla}, N. 2012, \bibinfo{title}{{Present-day cosmic
  abundances. A comprehensive study of nearby early B-type stars and
  implications for stellar and Galactic evolution and interstellar dust
  models},} \aap, 539, A143, \dodoi{10.1051/0004-6361/201118158}

% type= article
\bibitem[{G. {Pietrzy{\'n}ski} {et~al.}(2019){Pietrzy{\'n}ski}, {Graczyk},
  {Gallenne}, {Gieren}, {Thompson}, {Pilecki}, {Karczmarek}, {G{\'o}rski},
  {Suchomska}, {Taormina}, {Zgirski}, {Wielg{\'o}rski}, {Ko{\l}aczkowski},
  {Konorski}, {Villanova}, {Nardetto}, {Kervella}, {Bresolin}, {Kudritzki},
  {Storm}, {Smolec}, \& {Narloch}}]{Pietrzynski:2019}
{Pietrzy{\'n}ski}, G., {Graczyk}, D., {Gallenne}, A., {et~al.} 2019,
  \bibinfo{title}{{A distance to the Large Magellanic Cloud that is precise to
  one per cent},} \nat, 567, 200, \dodoi{10.1038/s41586-019-0999-4}

% type= article
\bibitem[{ {Planck Collaboration} {et~al.}(2020){Planck Collaboration},
  {Aghanim}, {Akrami}, {Ashdown}, {Aumont}, {Baccigalupi}, {Ballardini},
  {Banday}, {Barreiro}, {Bartolo}, {Basak}, {Battye}, {Benabed}, {Bernard},
  {Bersanelli}, {Bielewicz}, {Bock}, {Bond}, {Borrill}, {Bouchet}, {Boulanger},
  {Bucher}, {Burigana}, {Butler}, {Calabrese}, {Cardoso}, {Carron},
  {Challinor}, {Chiang}, {Chluba}, {Colombo}, {Combet}, {Contreras}, {Crill},
  {Cuttaia}, {de Bernardis}, {de Zotti}, {Delabrouille}, {Delouis}, {Di
  Valentino}, {Diego}, {Dor{\'e}}, {Douspis}, {Ducout}, {Dupac}, {Dusini},
  {Efstathiou}, {Elsner}, {En{\ss}lin}, {Eriksen}, {Fantaye}, {Farhang},
  {Fergusson}, {Fernandez-Cobos}, {Finelli}, {Forastieri}, {Frailis},
  {Fraisse}, {Franceschi}, {Frolov}, {Galeotta}, {Galli}, {Ganga},
  {G{\'e}nova-Santos}, {Gerbino}, {Ghosh}, {Gonz{\'a}lez-Nuevo}, {G{\'o}rski},
  {Gratton}, {Gruppuso}, {Gudmundsson}, {Hamann}, {Handley}, {Hansen},
  {Herranz}, {Hildebrandt}, {Hivon}, {Huang}, {Jaffe}, {Jones}, {Karakci},
  {Keih{\"a}nen}, {Keskitalo}, {Kiiveri}, {Kim}, {Kisner}, {Knox},
  {Krachmalnicoff}, {Kunz}, {Kurki-Suonio}, {Lagache}, {Lamarre}, {Lasenby},
  {Lattanzi}, {Lawrence}, {Le Jeune}, {Lemos}, {Lesgourgues}, {Levrier},
  {Lewis}, {Liguori}, {Lilje}, {Lilley}, {Lindholm}, {L{\'o}pez-Caniego},
  {Lubin}, {Ma}, {Mac{\'\i}as-P{\'e}rez}, {Maggio}, {Maino}, {Mandolesi},
  {Mangilli}, {Marcos-Caballero}, {Maris}, {Martin}, {Martinelli},
  {Mart{\'\i}nez-Gonz{\'a}lez}, {Matarrese}, {Mauri}, {McEwen}, {Meinhold},
  {Melchiorri}, {Mennella}, {Migliaccio}, {Millea}, {Mitra},
  {Miville-Desch{\^e}nes}, {Molinari}, {Montier}, {Morgante}, {Moss}, {Natoli},
  {N{\o}rgaard-Nielsen}, {Pagano}, {Paoletti}, {Partridge}, {Patanchon},
  {Peiris}, {Perrotta}, {Pettorino}, {Piacentini}, {Polastri}, {Polenta},
  {Puget}, {Rachen}, {Reinecke}, {Remazeilles}, {Renzi}, {Rocha}, {Rosset},
  {Roudier}, {Rubi{\~n}o-Mart{\'\i}n}, {Ruiz-Granados}, {Salvati}, {Sandri},
  {Savelainen}, {Scott}, {Shellard}, {Sirignano}, {Sirri}, {Spencer},
  {Sunyaev}, {Suur-Uski}, {Tauber}, {Tavagnacco}, {Tenti}, {Toffolatti},
  {Tomasi}, {Trombetti}, {Valenziano}, {Valiviita}, {Van Tent}, {Vibert},
  {Vielva}, {Villa}, {Vittorio}, {Wandelt}, {Wehus}, {White}, {White},
  {Zacchei}, \& {Zonca}}]{Planck:2020}
{Planck Collaboration}, {Aghanim}, N., {Akrami}, Y., {et~al.} 2020,
  \bibinfo{title}{{Planck 2018 results. VI. Cosmological parameters},} \aap,
  641, A6, \dodoi{10.1051/0004-6361/201833910}

% type= article
\bibitem[{N. {Przybilla} {et~al.}(2006){Przybilla}, {Butler}, {Becker}, \&
  {Kudritzki}}]{Przybilla:2006}
{Przybilla}, N., {Butler}, K., {Becker}, S.~R., \& {Kudritzki}, R.~P. 2006,
  \bibinfo{title}{{Quantitative spectroscopy of BA-type supergiants},} \aap,
  445, 1099, \dodoi{10.1051/0004-6361:20053832}

% type= article
\bibitem[{A.~G. {Riess} {et~al.}(2022){Riess}, {Yuan}, {Macri}, {Scolnic},
  {Brout}, {Casertano}, {Jones}, {Murakami}, {Anand}, {Breuval}, {Brink},
  {Filippenko}, {Hoffmann}, {Jha}, {D'arcy Kenworthy}, {Mackenty}, {Stahl}, \&
  {Zheng}}]{Riess:2022}
{Riess}, A.~G., {Yuan}, W., {Macri}, L.~M., {et~al.} 2022, \bibinfo{title}{{A
  Comprehensive Measurement of the Local Value of the Hubble Constant with 1 km
  s$^{-1}$ Mpc$^{-1}$ Uncertainty from the Hubble Space Telescope and the SH0ES
  Team},} \apjl, 934, L7, \dodoi{10.3847/2041-8213/ac5c5b}

% type= article
\bibitem[{A.~G. {Riess} {et~al.}(2024){Riess}, {Scolnic}, {Anand}, {Breuval},
  {Casertano}, {Macri}, {Li}, {Yuan}, {Huang}, {Jha}, {Murakami}, {Beaton},
  {Brout}, {Wu}, {Addison}, {Bennett}, {Anderson}, {Filippenko}, \&
  {Carr}}]{Riess:2024}
{Riess}, A.~G., {Scolnic}, D., {Anand}, G.~S., {et~al.} 2024,
  \bibinfo{title}{{JWST Validates HST Distance Measurements: Selection of
  Supernova Subsample Explains Differences in JWST Estimates of Local H
  $_{0}$},} \apj, 977, 120, \dodoi{10.3847/1538-4357/ad8c21}

% type= article
\bibitem[{N. {Sch{\"o}neberg} {et~al.}(2022){Sch{\"o}neberg}, {Abell{\'a}n},
  {S{\'a}nchez}, {Witte}, {Poulin}, \& {Lesgourgues}}]{Schoeneberg:2022}
{Sch{\"o}neberg}, N., {Abell{\'a}n}, G.~F., {S{\'a}nchez}, A.~P., {et~al.}
  2022, \bibinfo{title}{{The H$_{0}$ Olympics: A fair ranking of proposed
  models},} \physrep, 984, 1, \dodoi{10.1016/j.physrep.2022.07.001}

% type= article
\bibitem[{D. {Scolnic} {et~al.}(2024){Scolnic}, {Boubel}, {Byrne}, {Riess}, \&
  {Anand}}]{Scolnic2024}
{Scolnic}, D., {Boubel}, P., {Byrne}, J., {Riess}, A.~G., \& {Anand}, G.~S.
  2024, \bibinfo{title}{{Calibrating the Tully-Fisher Relation to Measure the
  Hubble Constant},} arXiv e-prints, arXiv:2412.08449,
  \dodoi{10.48550/arXiv.2412.08449}

% type= article
\bibitem[{E. {Sextl} {et~al.}(2021){Sextl}, {Kudritzki}, {Weller}, {Urbaneja},
  \& {Weiss}}]{Sextl:2021}
{Sextl}, E., {Kudritzki}, R.-P., {Weller}, J., {Urbaneja}, M.~A., \& {Weiss},
  A. 2021, \bibinfo{title}{{Modified Gravity and the Flux-weighted
  Gravity-Luminosity Relationship of Blue Supergiant Stars},} \apj, 914, 94,
  \dodoi{10.3847/1538-4357/abfafa}

% type= article
\bibitem[{R.~B. {Tully} \& H.~M. {Courtois}(2012){Tully} \&
  {Courtois}}]{Tully:2012}
{Tully}, R.~B., \& {Courtois}, H.~M. 2012, \bibinfo{title}{{Cosmicflows-2:
  I-band Luminosity-H I Linewidth Calibration},} \apj, 749, 78,
  \dodoi{10.1088/0004-637X/749/1/78}

% type= article
\bibitem[{R.~B. {Tully} \& J.~R. {Fisher}(1977){Tully} \&
  {Fisher}}]{Tully:1977}
{Tully}, R.~B., \& {Fisher}, J.~R. 1977, \bibinfo{title}{{A new method of
  determining distances to galaxies.},} \aap, 54, 661

% type= article
\bibitem[{M.~A. {Urbaneja} {et~al.}(2017){Urbaneja}, {Kudritzki}, {Gieren},
  {Pietrzy{\'n}ski}, {Bresolin}, \& {Przybilla}}]{Urbaneja:2017}
{Urbaneja}, M.~A., {Kudritzki}, R.-P., {Gieren}, W., {et~al.} 2017,
  \bibinfo{title}{{LMC Blue Supergiant Stars and the Calibration of the
  Flux-weighted Gravity-Luminosity Relationship},} \aj, 154, 102,
  \dodoi{10.3847/1538-3881/aa79a8}

\end{thebibliography}
\bibliographystyle{aasjournalv7}

\end{document}